\documentclass[aps,prc,preprint,groupedaddress,showpacs,showkeys]{revtex4}
\usepackage{graphicx}
\usepackage{dcolumn}
\usepackage{bm}
\usepackage{epsfig}
\begin{document}
%
\bibliographystyle{apsrev}
%
%
%
\title{Hypernuclear Spectroscopy Using the $(e,e^{\prime}
\mbox{K}^{+})$ Reaction}
%

%
%
\author{
L. Yuan$^{a}$,
M. Sarsour$^{b}$,
T. Miyoshi$^{c}$,
X. Zhu$^{a}$,
A. Ahmidouch$^{d}$, 
D. Androic$^{e}$, 
T. Angelescu$^{f}$, 
R. Asaturyan$^{g}$, 
S. Avery$^{a}$, 
O.K. Baker$^{a,i}$, 
I. Bertovic$^{e}$,
H. Breuer$^{h}$,
R. Carlini$^{i}$, 
J. Cha$^{a}$, 
R. Chrien$^{j}$, 
M. Christy$^{a}$, 
L. Cole$^{a}$, 
S. Danagoulian$^{d}$, 
D. Dehnhard$^{k}$, 
M. Elaasar$^{l}$, 
A. Empl$^{b}$, 
R. Ent$^{i}$, 
H. Fenker$^{i}$, 
Y. Fujii$^{c}$, 
M. Furic$^{e}$, 
L. Gan$^{a}$, 
K. Garrow$^{i}$, 
A. Gasparian$^{a}$, 
P. Gueye$^{a}$, 
M. Harvey$^{a}$, 
O. Hashimoto$^{c}$, 
W. Hinton$^{a}$, 
B. Hu$^{a}$, 
E. Hungerford$^{b}$, 
C. Jackson$^{a}$, 
K. Johnston$^{n}$, 
H. Juengst$^{k}$, 
C. Keppel$^{a}$,
K. Lan$^{b}$, 
Y. Liang$^{a}$, 
V.P. Likhachev$^{o}$, 
J.H. Liu$^{k}$, 
D. Mack$^{i}$, 
A. Margaryan$^{g}$, 
P. Markowitz$^{p}$, 
H. Mkrtchyan$^{g}$, 
S. N. Nakamura$^{c}$
T. Petkovic$^{e}$,
J. Reinhold$^{p}$, 
J. Roche$^{q}$,
Y. Sato$^{c,a}$, 
R. Sawafta$^{d}$, 
N. Simicevic$^{n}$, 
G. Smith$^{i}$, 
S. Stepanyan$^{g}$, 
V. Tadevosyan$^{g}$, 
T. Takahashi$^{c}$, 
K. Tanida$^{r}$,
L. Tang$^{a,i}$, 
M. Ukai$^{c}$, 
A. Uzzle$^{a}$, 
W. Vulcan$^{i}$, 
S. Wells$^{n}$, 
S. Wood$^{i}$, 
G. Xu$^{b}$, 
H. Yamaguchi$^{c}$, 
C. Yan$^{i}$\\
(HNSS Collaboration)}
%
\affiliation{
$^{a}$Hampton University, Hampton, VA 23668;
$^{b}$University of Houston, Houston, TX 77204;
$^{c}$Tohoku University, Sendai, 980-8578, Japan; 
$^{d}$North Carolina A\&T State University, Greensboro, NC 27411;
$^{e}$University of Zagreb, Zagreb, Croatia;
$^{f}$University of Bucharest, Bucharest, Romania;
$^{g}$Yerevan Physics Institute, Yerevan, Armenia;
$^{h}$University of Maryland, College Park, MD 20742;
$^{i}$Thomas Jefferson National Accelerator Facility, 
Newport News, VA 23606;
$^{j}$Brookhaven National Laboratory, Upton, NY 11973;
$^{k}$University of Minnesota, Minneapolis, MN 55455; 
$^{l}$Southern University at New Orleans, New Orleans, LA 70126;
$^{m}$Rensselaer Polytechnic Institute, Troy, NY 12180;
$^{n}$Louisiana Tech University, Ruston, LA 71272; 
$^{o}$University of Sao Paulo, Sao Paulo, Brazil; 
$^{p}$Florida International University, Miami, FL 33199;
$^{q}$College of William and Mary, Williamsburg, VA 23187; 
$^{r}$University of Tokyo, Tokyo 113-0033, Japan
}
%
\date{\today}
%
%
\begin{abstract}

A pioneering experiment in $\Lambda$
hypernuclear spectroscopy, undertaken at the Thomas Jefferson National
Accelerator Facility (Jlab), was recently reported.
The experiment used the high- 
precision, continuous electron beam at Jlab,
and a special arrangement of spectrometer 
magnets to measure the spectrum from $^{nat}$C
and $^{7}$Li targets using the $(e,e^{\prime} K^{+})$ reaction. 
The $^{12}_{\Lambda}$B spectrum 
was previously published.  This experiment is 
now reported in more detail, with
improved results for the $^{12}_{\Lambda}$B spectrum.  In
addition, unpublished results of the $^{7}_{\Lambda}$He 
spectrum are also shown. This later spectrum 
indicates the need for a more
detailed few-body calculation of the hypernucleus and 
the reaction process. The
success of this experiment demonstrates that the $(e,e^{\prime}K^{+})$
reaction can be effectively used as a high resolution tool to study
hypernuclear spectra, ant its use should be vigorously pursued.

\end{abstract}

\maketitle

%
%
\section{Introduction}

The introduction of strangeness into the nuclear medium 
challenges conventional models of this low-energy, hadronic, many-body system.
Of particular interest is the fact that one pion
exchange (OPE) between a $\Lambda$ and a nucleon does not occur due
to conservation of isospin.
Thus higher mass meson exchanges, including the 
two-pion exchange coupling of a
lambda to a nucleon through an intermediate sigma ($\Lambda$N
$\rightarrow$ $\Sigma$N $\rightarrow$ $\Lambda$N),
is significant, and leads to sizable charge asymmetry
and three-body forces\cite{gibson}.  In addition,
the strangeness degree of freedom allows the nucleus to rearrange
by taking advantage of SU(3) flavor symmetry, in order
to maximize the nuclear binding energy\cite{millener1}. 

The $\Lambda$ can also be used as a probe of the nuclear medium.  
If the $\Lambda$ is
considered a fundamental particle and remains identifiable as such
within the nucleus,  a hypernuclear
$\Lambda$ will sample the nuclear interior.  Thus various features
such as $\Lambda$ decay and
$\Lambda$N interactions in heavier hypernuclear systems can be
extremely interesting.  The hypernuclear system can
therefore better illuminate
features which would be more obscured in conventional nuclear
systems. 

Also, as it is essentially impossible to directly determine the
elementary $\Lambda$-Nucleon potential, the $\Lambda$N interaction
can presently only be extracted from 
hypernuclei. Fortunately, the $\Lambda$N
interaction is weak, so that one can with
some confidence, obtain a $\Lambda$N potential after the effective
$\Lambda$-Nucleus interaction is found from an analysis of
hypernuclear spectra\cite{millener2}.  Such information illuminates 
the SU(3)$_{flavor}$ baryon-baryon interaction at normal nuclear
densities.  This information can then serve as a 
normalization point, to
extrapolate the interaction to matter-densities found in 
neutron stars, where mixtures of nucleons and hyperons could form a
stable system\cite{prakash}.  
  
Traditionally, hypernuclei have been produced with secondary beams of
kaons or pions, as shown in Figure 1a.  Because the $(K^{-},\pi^{-})$
reaction is exothermic, the momentum transfer to the $\Lambda$ can be
chosen to be small.  In this situation the cross section 
to substitution states ( {\it i.e} states where
the $\Lambda$ acquires the same shell quantum numbers as those of the
neutron which it replaces) is relatively large.  On the other hand, the
$(\pi^{+},K^{+})$ reaction has momentum transfers comparable to the
nuclear Fermi-momentum, and the cross section preferentially populates
states with high angular momentum transfers \cite{milner,bando}. Neither of 
these two reactions has significant spin-flip amplitude at forward angles
where the cross sections are experimentally accessible.  Thus all
these spectra are dominated by transitions to natural parity states.

\begin{figure}[htb!]
\begin{center}
\epsfxsize = 15cm
\epsffile{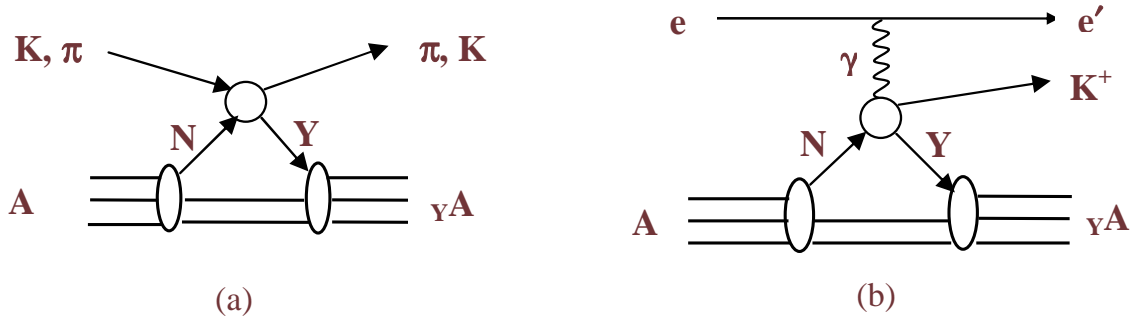}
\caption{A schematic representation of the a) mesonic and b)
  electromagnetic production processes}
\label{fig:production}
\end{center}
\end{figure}

Aside from early emulsion experiments, mesonic production of
hypernuclei has generally provided hypernuclear 
spectra with energy resolutions $\ge$ 2 MeV.  This is due to the
intrinsic resolutions
of secondary mesonic beamlines, and the target thicknesses required to
obtain sufficient counting rates.  One previous study did achieve a spectrum
resolution of approximately 1.5 MeV for the $^{12}_\Lambda$C
hypernucleus, using a thin 
target and devoting substantial time to data collection\cite{hasegawa}.
This work demonstrated the importance of good resolution, as
significantly new information, showing some of the fine structure in the
$^{12}_{\Lambda}$C spectrum, was obtained.

Although, specific hypernuclear states below nucleon emission
threshold can be located within a 
few keV by detecting
de-excitation gammas\cite{tamura,tanida} in 
coincidence with a hypernuclear production
reaction, such experiments become more difficult in heavier systems
due to the number of transitions which must be unambiguously assigned
in an unknown spectrum.  It should be noted however, that
resolutions of a few hundred keV are  also
sufficient for many studies, since reaction selectivity and angular
dependence potentially allows extraction of the spectroscopic
factors to specific states\cite{rosenthal}.  Of course, a
reaction also provides a full spectrum of states which
can be clearly identified with a specific hypernucleus.
Indeed the excitation strength of the spectrum is of interest, as
it directly relates to the model that the reaction proceeds
through the interaction of the incident projectile
with an identifiable nucleon within the nuclear medium. 
Thus as an example apropos to the experiment reported here, if the
theoretical spectrum does not reproduce the experimental one, 
it is possible that
propagator re-normalization within the medium could be significant
\cite{cotanch}, requiring a modification of the single particle
picture of the reaction.

Electroproduction of hypernuclei is illustrated by Figure 1b.
Electroproduction traditionally has been used for precision studies
of nuclear structure, as the exchange of a colorless photon can be
accurately described by a first order perturbation calculation.
In addition, electron beams have excellent spatial and energy
resolutions.
Previously, electron accelerators had poor duty factors, significantly
impairing high singles rate, coincidence experiments.
 However, modern, continuous beam accelerators have now overcome this
limitation, and although the cross section for kaon electroproduction
is some 2 orders of magnitude smaller than hadronic reactions, this
can be compensated by increased beam intensity.  Targets can be
physically small and thin (10-50 mg/cm$^{2}$), allowing studies of
almost any isotope. The potential result for $(e,e^{\prime}K^{+})$
experiments, is an energy resolution of a few 
hundred keV with reasonable
counting rates up to at least medium weight hypernuclei\cite{hungerford}.

The $(e,e^{\prime}K^{+})$ reaction, because of the absorption of the
spin 1 virtual photon, has high spin-flip probability even at forward
angles. In addition the momentum transfer is high, approximately 300
MeV/c at zero degrees for 1500 MeV incident photons, so the
resulting reaction is expected to predominantly excite spin-flip
transitions to  
spin-stretched states of unnatural parity\cite{motoba}.   These states are not
strongly excited in hadronic production, and the electromagnetic process
acts on a proton rather than a neutron creating proton-hole,
$\Lambda$-particle states, charge symmetric to those previously studied with
meson beams.
Precision experiments, comparing mirror hypernuclei, are needed in
fact, to extract
the charge asymmetry in the $\Lambda$N potential. 

An initial experiment\cite{miyoshi1},
in Hall C at Thomas Jefferson National Acceleration Facility (Jlab)
has been previously reported.
The unique features of this experiment include;
\begin{enumerate}
\item  resolutions of less than 1 MeV FWHM;
\item spectra using the $(e,e^{\prime}K^{+})$ 
reaction for the first time;
\item spectra which show unnatural parity
(spin flip) states;
\item isospin mirrored spectra compared to those produced
by the (K$^{-},\pi^{-}$) and ($\pi^{+}$,K$^{+}$) reactions;
\item thin targets and low luminosities; and
\item a high-rate silicon strip detector mounted in the focal 
plane of the electron spectrometer.
\end{enumerate}
This paper discusses the experiment in more detail, and presents
an improved spectrum of the $^{12}_{\Lambda}$B hypernucleus as well as
a previously unpublished spectrum of the 
$^{7}$Li$(e,e^{\prime}K^{+})^{7}_{\Lambda}$He reaction.

\section{ Experimental Details}

In electroproduction, the $\Lambda$ and $K^{+}$ particles are created 
associatively via an interaction between a virtual photon and a proton
in the nucleus, $p(\gamma,K^{+})\Lambda$. The hypernucleus, $_{\Lambda}A$,  
is formed by coupling this $\Lambda$ to the residual nuclear
core, (Z-1), as shown in Figure 1(b).  In electroproduction, the 
energy and momentum of the virtual photon are defined by $\omega =
E(e)-E(e^{\prime})$ and $\vec{q} = \vec{p}(e) - \vec{p}(e^{\prime})$,
respectively.  The four-momentum transfer of the electron is then
given by $Q^{2} = q^{2}- \omega^{2}$. 
Since the elementary cross section, and the nuclear form
factor, fall rapidly with increasing $Q^{2}$, experiments must be done
within a small angular range around the direction of the virtual photon.
In addition as discussed below, the virtual photon flux is
maximized for an electron scattering angle near zero
degrees. Thus the experimental geometry requires two spectrometer
arms, one to detect the scattered electron and one to detect the kaon, 
both placed at extremely forward angles.

The electroproduction cross section can be
expressed \cite{donnelly}  by; \\ 
 
\begin{math}
\noindent \hspace*{0.5cm}\frac{\displaystyle\partial^{3}\sigma}
{\displaystyle\partial\mbox{E'}_{e}\partial\Omega'_{e} \partial\Omega_{k}} =
\Gamma\left[\frac{\displaystyle\partial
\sigma_{T}}{\displaystyle\partial\Omega_{k}} + 
\epsilon\frac{\displaystyle\partial\sigma_{l}}
{\displaystyle\partial\Omega_{k}} +
\epsilon \mbox{cos(2}\phi) +
\mbox{cos}(\phi_{k})\sqrt{2\epsilon(1+\epsilon)}
\frac{\displaystyle\partial\sigma_{I}}{\displaystyle\partial\Omega_{k}}\right].\\
\end{math} 

\noindent The factor, $\Gamma$, is the virtual flux factor evaluated with
electron kinematics in the lab frame.  It has the form;\\

\begin{math}
\noindent \hspace*{0.5cm}\Gamma =
\frac{\displaystyle\alpha}{\displaystyle 2\pi^{2}\mbox{Q}^{2}}
\frac{\displaystyle\mbox{E}_{\gamma}}
{\displaystyle 1-\epsilon}\frac{\displaystyle\mbox{E}^{\prime}_{e}}
{\mbox{E}_{e}}.\\
\end{math}

\noindent In the above equation, $\epsilon$ is the polarization
factor; \\

\begin{math}
\noindent \hspace*{0.5cm}\epsilon = \left[ 1 +
\frac{\displaystyle 2|{\bf k}|^{2}}{\displaystyle 
\mbox{Q}^{2}}\mbox{tan}^{2}(\Theta_{e}/2)\right]^{-1}.\\
\end{math}

\noindent For those virtual photons almost on the mass shell,
$\mbox{Q}^{2}={\bf \mbox{p}}^{2}_{\gamma}-\mbox{E}^{2}_{\gamma} 
\rightarrow 0$.   The label on each of the
cross section expressions ($T, L, P,$ and $I$) represent
transverse, longitudinal, polarization, and interference terms.
For real photons of course, Q$^{2}\rightarrow$0, so only the transverse
cross section is non-vanishing, and for the experimental geometry
used here, the cross section is completely dominated by the transverse
component.  Thus the electroproduction cross section may be replaced,
to good approximation, by the photoproduction value times the flux factor.

Experimentally, $\Gamma$ is integrated
over the  angular and momentum acceptances of the electron spectrometer.
In order to maximize the cross section of the elementary, 
$p(\gamma,k^{+})\Lambda$ reaction, the
photon energy is chosen to be about 1.5 GeV.  In addition, to keep
strangeness production limited essentially to kaons and $\Lambda$s,
the energy, E(e), of the incident 
electron is set to be $\approx \, 1.8$ GeV. In this way, backgrounds from
unwanted reactions are reduced.  This also allows a physically small, 
low-momentum electron spectrometer to be employed, as the 
scattered electron energy,$(E_e^{\prime})$, is about 0.3 GeV. 

\begin{figure}[htb!]
\begin{center}
\epsfxsize = 15cm
\epsffile{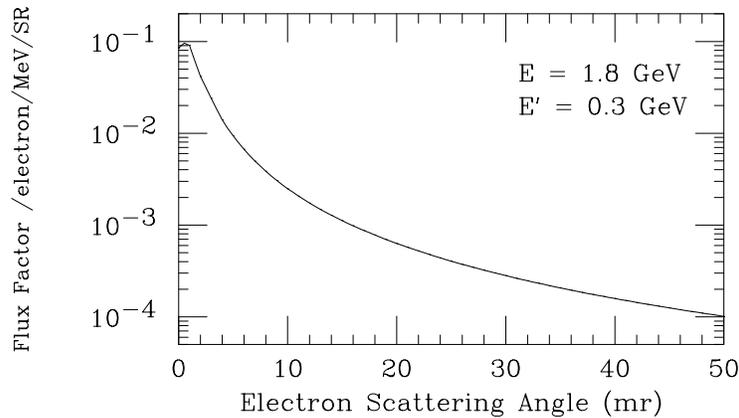}
\caption{The virtual photon flux factor as a function of the electron
  scattering angle}
\label{fig:flux-factor}
\end{center}
\end{figure}

Figure \ref{fig:flux-factor}. shows the calculated virtual photon
flux factor in units of photons per  
electron per MeV-sr for the chosen kinematics. This flux factor
peaks at zero  
degrees and falls rapidly as the scattering angle increases\cite{xu,hyde}. With
electrons  detected at zero degrees, a large 
percentage of the scattered electrons are captured by even a small
solid angle, increasing the coincidence
probability between these electrons and the reaction kaons of interest. In
addition, because of the small beam spot ($\approx \, 100 \mu$m), the
$\approx \, 0^{\circ}$ electron scattering angle, and the small
momentum value of the scattered electron, it is sufficient to only measure the
electron position on the spectrometer focal plane to ensure
excellent energy resolution.  However, the disadvantage
of this geometry is a high electron background rate from target
bremsstrahlung, which ultimately limits the usable beam luminosity.
  
\begin{figure}[htb!]
\begin{center}
\epsfxsize = 9cm
\epsffile{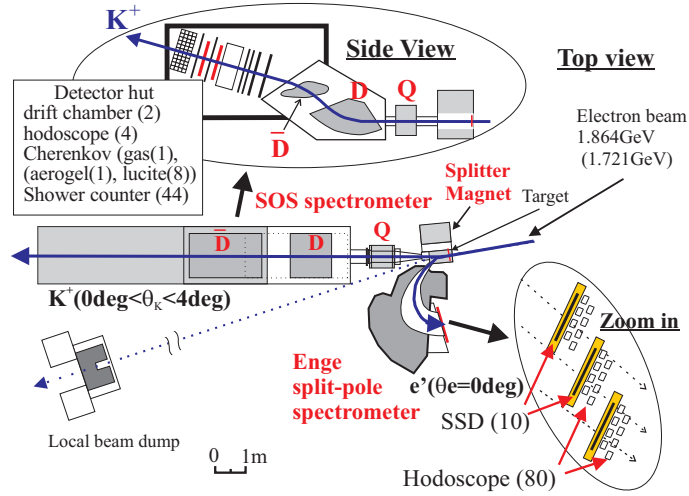}
\caption{The experimental plan view showing both the kaon
    spectrometer (SOS) and the electron spectrometer (ENGE).  The SOS
is a QD\={D} spectrometer with Q an entrance quadrupole, and D\={D} two
dipoles bending in opposite directions, providing large momentum
acceptance but reducing dispersion.}
\label{fig:layout}
\end{center}
\end{figure}

Once the choice of the incident and scattered electron momenta are fixed, the
reaction kaon momenta are determined by the kaon reaction angle. In
this experiment, the chosen kinematics produced a
kaon momenta of $\approx \, 1.2 GeV/c$,
providing a 3-momentum transfer of $\approx \, 300 MeV/c$ to the
recoiling $\Lambda$.  The kaon momentum provides a
reasonable kaon survival fraction, and allows $\pi/K$ discrimination using
threshold aerogel Cerenkov detectors coupled with time of
flight. Figure \ref{fig:layout} shows
a schematic view of the experimental layout.

\subsection{The Beam}

The beam has a bunch width of 1.67 ps with a bunch separation of
2 ns. While the absolute value of the energy of the incident
electrons was unimportant (although for kinematic reasons to be
discussed below, it did need to be determined), it
was extremely important to precisely maintain whatever this energy was
over the several 
weeks of the experiment.  Thus
the beam momentum is locked by a Fast Feedback 
Energy Lock System installed in an arc of the accelerator. This system 
measured, at a repetition rate of 1 khz, the beam parameters at the
entrance, the position of maximum momentum dispersion, and the exit
of the arc, to extract an energy correction factor.  This correction 
was then applied to the last cavity of the accelerator, maintaining a
constant beam energy.  The
feedback lock controlled the total energy of the primary electron beam to a
$\delta p/p \, \le \, 10^{-4}$.  The intrinsic energy spread in the
beam was controlled by a tune of the injector.
 A similar lock system maintained
the beam position on target within 100 $\mu$m.  Although the 
intrinsic spot size
was tuned to be $\le \, 100 \, \mu$m,
the beam was de-focused to $4 \times 4$ $mm^{2}$ by a fast raster 
when incident on the
CH$_{2}$ target,  to reduce beam heating.

Beam intensities were set to produce an acceptable
signal to accidental ratio, which for the C target was
approximately 0.6 $\mu$A, or an experimental luminosity of
approximately $4 \times 10^{33}$ cm$^{-2}$-s$^{-1}$. Due to the lower
radiation length of the $^{7}$Li  
target, a higher beam current of about 0.8 $\mu$A was used.  To
protect the CH$_{2}$ target, the beam current was kept below
1.5 $\mu$A.  

Finally, to satisfy conflicting 
beam requirements when simultaneously operating several experiments in
the different 
experimental Halls, data had to be acquired at two time periods using 
two different beam energies, 1721 and 1864 MeV. These different energy
data were analyzed separately, but the kinematical conditions
for the two beam  
energies were close, and the spectra showed no energy dependence within
statistics.  Therefore these data were 
summed after separate analysis to increase the statistical significance of
the final spectrum. In addition to
calibrations, about 400 hours of data were collected with the
$^{nat}$C target and 
about 120 hours with the $^{7}$Li target.

\subsection{ The Splitting Magnet and Targets}

In order to detect both scattered electrons and positively charged kaons near 
zero degrees, a ``C'' magnetic dipole (splitter) was used.  The
target was positioned at the upstream side of the effective field
boundary of this magnet. The splitter respectively deflected electrons 
scattered at approximately 0 degrees and kaons at approximately 
2 degrees by 33 and 16 degrees, respectively.

 Three different, target foils were employed, CH$_{2}$, 8.8 mg/cm$^{2}$, 
$^{nat}$C, 22 mg/cm$^{2}$, and
 $^{7}$Li, 19 mg/cm$^{2}$. 
By observing the $p(e,e^{\prime}K^{+})\Lambda$ and
 $p(e,e^{\prime}K^{+})\Sigma$, the hydrogen in the 
 CH$_{2}$ foil was used for energy calibration and
 optimization  of the spectrometer optics. These procedures are
 described in the subsections below. 

\section{The Kaon Spectrometer}

A short orbit spectrometer (SOS) is one of two existing magnetic
spectrometers in Hall C at Jlab, and as it has a flight path of
$\approx \, 10$m, this spectrometer is particularly
useful for the detection of particles with short half-lives.  However, it
has low dispersion and large momentum acceptance, and these characteristics 
are not well matched to
the present experimental geometry.  Still the SOS was 
mounted at the Hall C pivot and had the sophisticated particle
identification (PID) package\cite{niculescu} required to identify
kaons within the 
large background of pions and positrons, so it was chosen as the kaon
spectrometer for this first $(e,e^{\prime}K^{+})$ experiment.  It was
expected that the
overall resolution would be dominated by this spectrometer\cite{tang}.

The solid angle acceptance of the splitter/SOS spectrometer system was
approximately 5 msr, covering a range of scattering angles from 0 to 4 
deg.  The error in the reconstructed scattering angle was about 13 mr (FWHM),
and was dominated by the horizontal angular error measurement.
This contributed about 200 keV to the missing mass 
resolution when the recoil atomic number was $\ge \, 6$. The central momentum
of the SOS was set to 1.2 GeV/c.  The acceptance was $\approx \,
46$\%, but only the central $\pm \, 15$\%  was 
useful.  This acceptance was nearly flat within the missing mass range
of interest. 

The standard SOS detector package was used. It consisted of: 1) 
two sets of tracking chambers separated by 0.5 meters;
2) four scintillation hodoscope planes; 3) one aerogel Cerenkov (AC)
counter with an optical index of 1.03; 4) one Lucite, total internally 
reflecting Cerenkov (LC) counter 
with index 1.49; 5) one gas Cerenkov (GC) detector; and 6) 
3 layers of lead-glass shower counters.  The tracking detectors were
used to determine the position and angle of the particle on the focal plane, and
by projection, its scattering angle from the target.  The scintillator 
hodoscopes were used to
localize tracks in the wire chambers, and to obtain
timing and time of flight (TOF) information for PID.  The aerogel
Cerenkov detector was used to veto pions and positrons, and the Lucite
counter was used to remove protons by tagging high-beta particles. The
gas Cerenkov detector and the lead-glass shower counters were used to
remove positrons.  

\subsection{The Electron Spectrometer}

The scattered electrons were detected in a split-pole, magnetic
spectrometer\cite{spencer},  
which was well matched to the geometrical kinematics and acceptances.  The
spectrometer coupled with the phase space of the incident beam, had 
the capability of obtaining $5 \times 10^{-4}$ resolution (FWHM $\delta
p/p$).  The central momentum was chosen to be 300 MeV/c 
with a momentum acceptance of $\approx \, 120$ MeV/c. The solid angle 
acceptance of the combined splitter/split-pole system was about 9 msr,
which effectively tagged about 35\% of the virtual photon flux. This
provided a spread in the photon flux momentum of $\approx \, 120$ MeV/c,
centered around $\approx \, 1500$ MeV/c.  In summary, the geometry of
the electron arm was
possible because of the excellent phase space of the incident electron
beam, the thin targets which limited multiple scattering, and the
extremely forward peaking of the virtual photon flux factor.

However, target bremsstrahlung also peaks at zero degrees and large numbers
of scattered electrons are expected to enter the split-pole
spectrometer\cite{xu} . 
In fact the experimental luminosity was set by accepting a
total rate of $\approx \, 2 \times 10^{8}$ $s^{1}$ on the 
instrumented portion of the
focal plane. To operate at this rate, the detection system required that only
the focal plane position of a scattered electron was needed to obtain
the required 
resolution, since more detailed tracking would have been very
inefficient.  The focal plane 
detector\cite{miyoshi2} was composed of 10, one-dimensional 
silicon strips segments 
(SSD), each having 144 strips with a pitch of 0.5mm.  These segments were
placed approximately perpendicular to the electrons,
which were incident at $\approx \, 47^{\circ}$ on a 72 cm length of the
the focal plane.  The singles rate per strip was on average 
$\approx \, 10^{5} \, s^{-1}$.

A set of 8 scintillation strip counters in a hodoscope arrangement
were positioned directly behind each of the SSD segments. These strips
were 1cm wide, 6cm long and 0.4 cm thick, viewed at one end through a
light guide 
by a 3469 Hamamatsu photomultiplier.  Rates per scintillator were
found to be $\le \ 1.5 \times 10^{6}$ hz, and no change in time
resolution was observed up to rates $\le \ 2.5 \times 10^{6}$ hz. The SSD
provided the position of an electron event to within 500 $\mu$m and
the scintillation hodoscope provided event timing to 250 ps
($\sigma$).

\subsection{Pion/Kaon Discrimination}

It was expected that the numbers of positrons, pions and protons in
the SOS would be very much larger than the number of kaons.  Indeed
the measured flux per second of positrons, pions, protons, and kaons 
from the C target, was $ 10^{5}$, $1.4 \times 10^{3}$, 140, and 0.4,
respectively. Therefore excellent particle identification was
required, not only in the analysis, but also in the hardware trigger.

The standard SOS detector package was used to identify
kaons, and its description and operation have been previously
discussed\cite{niculescu}.   The large flux of positrons 
was due to the acceptance of scattering angles down to zero degrees, 
where positrons from Dalitz pairs, created in the target,
were observed.  Positrons were easily identified and could have been
removed in the trigger by the lead-glass shower counter, but
detection of the Dalitz pairs provided a useful confirmation 
of the experimental resolution.  

\begin{figure}[htb!]
\begin{center}
\epsfxsize = 7cm
\epsffile{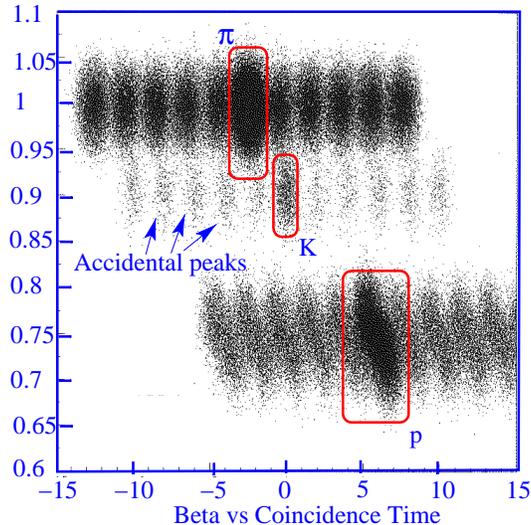}
\caption{A plot of SOS particle velocity vs the electron-kaon
  time-of-flight showing the separation of pions, kaons and protons}
\label{fig:pid}
\end{center}
\end{figure}

The coincident time resolution between an electron and a kaon was 230 (FWHM)
ps, after pulse height and path
length corrections were applied.  Coupled with a measure of the kaon
time-of-flight, 
the system time resolution was sufficient to identify the 
real and accidental coincidence peaks, and the true kaon coincident
events were  
selected by a two-dimensional cut on a real coincidence window of 2 ns
as shown in Figure \ref{fig:pid}.
Events selected from an average of eight nearby accidental coincidence
windows were 
used to obtain the shape and magnitude of the accidental background spectrum.

\subsection{The Expected System Resolution}

Since the entire beamline/spectrometer system was under vacuum, 
multiple scattering in the air and vacuum windows occurred only 
at the exit 
of the spectrometers.  Vacuum windows were located immediately before the
first tracking detectors so that this effect on the measured
track-position was minimized. Table 1 lists the expected 
contributions to the energy resolution.  As discussed above, the
contribution from the SOS spectrometer was expected to dominate.

\begin{table}[htb]
\begin{center}
\caption{ Contributions to the System Energy
  Resolution} 
\begin{tabular}{ccc}
\hline\hline
Source & Contribution & Resolution (keV) \\
\hline
Beam Energy & $10^{-4}$   &  $\le$ 180 \\
SOS momentum & $5.5\times 10^{-4}$   &  $\approx$ 660  \\
Split-pole   & $5\times 10^{-4}$  & 150 \\
Kaon Scattering Angle ($^{12}C$) & 13 mr & $\approx$ 200 \\
Target Energy Loss ($^{12}C$) & 1.7 keV/mg/cm$^{2}$ & 38  \\
Total  & &  $\approx$ 757 \\
\hline\hline
\end{tabular}
\label{trig_det}
\end{center}
\end{table}

\begin{figure}[htb!]
\begin{center}
\epsfxsize = 8cm
\epsffile{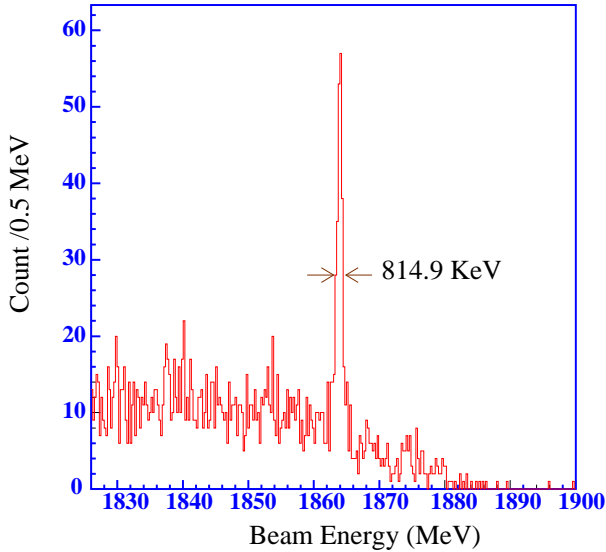}
\caption{Reconstructed beam energy from the measurement of Dalitz
  pairs detected by the spectrometer system.  As described in the
  text, this was used to confirm the energy resolution of the experiment.}
\label{fig:dalitz-pairs}
\end{center}
\end{figure}

The system resolution could be experimentally checked using Dalitz 
pairs from the $A(e,e^{\prime};(e^{+}e^{-}))A$  reaction, where both electrons 
were detected in the electron spectrometer and the positron in the
kaon spectrometer.  As the electrons and positrons are emitted essentially 
at zero degrees and the electron mass is negligible, 
the sum of the separate energies of these particles is 
equal to the beam energy. The width of the reconstructed beam-energy peak 
is about 815 keV (FWHM).  Unfortunately the kinematics limited the electron 
pair to the upper region of the momentum acceptance, where the resolution 
is maximized.  From an optical study of the resolution over the entire
focal plane, we estimate that the reconstructed Dalitz pair should be 
approximately 900 keV (FWHM).  This width provides an estimate of 
the experimental resolution which is due to the detection of only two
rather than three particles, but over the full momentum acceptance.
The experimental Dalitz pair spectrum is shown in Figure \ref{fig:dalitz-pairs}.

\subsection{  Rates and Background}

The singles rate in the electron arm was set to about $2 \times
10^{8}$ s$^{-1}$. As discussed previously, this rate was primarily 
a result of bremsstrahlung electrons.  Therefore, the experimental 
trigger was the much less frequent identification of a kaon in the 
kaon spectrometer. The coincidence spectra were then obtained later 
in off-line analysis.  The positrons from $e^{+}/e^{-}$ pair production
dominated the rate in the kaon spectrometer. These Dalitz pairs were
produced essentially at 0 degrees, and since the SOS acceptance covered 
0 to 4 degrees, they were accepted into the spectrometer.  However, the
combined use of vetoes from AC, GC, and shower counters substantially 
reduced events triggered by positrons, and they were completely 
eliminated in the off-line analysis. Rates from protons and pions were 
also reduced to approximately 1 kHz after on-line cuts by the AC and 
LC detectors.  The remaining protons and pions were also eliminated 
in the off-line analysis.

About 95\% of the background observed in the raw spectra was due to 
accidental coincidences. An evaluation of this background began by 
obtaining a spectrum of the time difference between the emission of the 
kaon and electron(s) in an event.  This time spectrum contained both 
the real, and a number of accidental peaks, separated by the 2ns time 
structure in the beam.  The summation of the analyzed spectra from 8 
of the accidental time-peaks provided a higher statistics measurement 
of the accidental background.  The remaining background (5\%) was due 
to real coincidences of pions (misidentified kaons) with electrons.  
The TOF path length separation between pions and kaons was about 2 ns, 
and therefore real coincidences could occur between electrons and
misidentified pions which were emitted 2 ns later but arrived in the 
time window of the real kaon peak.  The shape of the pion background 
in the missing-mass spectra was obtained by cutting on pions in the PID 
and analyzing coincident pions assuming $(e,e^{\prime}K^{+})$ kinematics.  
The absolute magnitude of this background was then obtained by normalizing 
the pion spectrum to the number  of background events.  

In addition, the time resolution of $\sigma \,\approx \, 230$ ps,
allowed the tails of the real kaon coincidence peak to overlap with
an accidental neighbor. Thus to calculate the cross section, the number 
of true coincident kaons lost from the real time peak was compensated 
by a cut-efficiency factor. 

\begin{figure}[htb!]
\begin{center}
\epsfxsize = 8cm
\epsffile{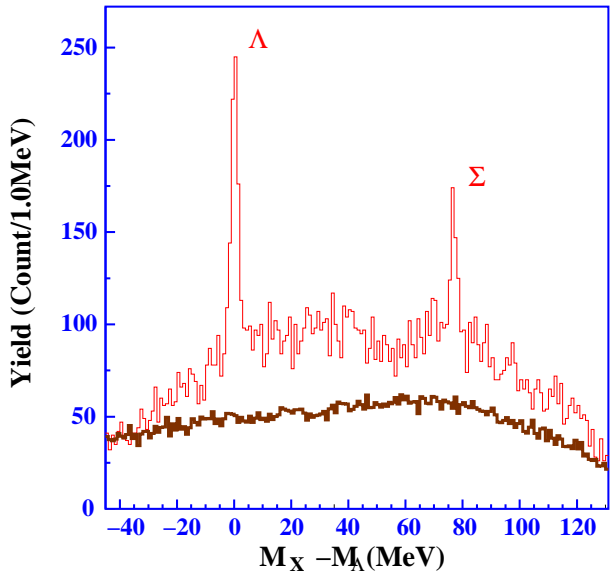}
\caption{The missing mass spectrum obtained 
from a CH$_{x}$ target showing both $\Lambda$ and $\Sigma$ production
from hydrogen in the target.  The solid histogram is the accidental
background.} 
\label{fig:cal}
\end{center}
\end{figure}

\subsection{  Calibrations, Spectrometer Optics, and Kinematics}

Analysis of the experiment required knowledge of the magnetic
transport coefficients of the spectrometers.
Although the coefficients for the SOS 
transport had been previously established in several experiments, 
the addition of the splitting magnet to the system required that 
they be re-determined.  For example, the angular acceptance of the 
SOS, normally 7 msr for a point target, was reduced to 5 msr by the 
splitter.

The reconstructed missing mass of a hypernuclear state is a 
function of the beam energy, the momenta
of the scattered electron and kaon, and the scattering angles. In a
two-dimensional space defined by the electron and kaon momenta, the
recoil missing mass is obtained by a projection of the events 
onto a locus line.  Using an
incorrect value of the beam energy or central momentum for
either spectrometer arm,
results in an incorrect position and slope of this locus line.
Therefore there is both an
incorrect kinematic position and width for various missing masses.
Thus we not only need the relative
values of scattering angles and the fractional change of the scattered
momentum with respect to central momenta, but also the absolute
values of the beam energy and central momenta of the spectrometers, or
the absolute values of the scattering angles of the coincident particles.
We have now obtained better calibration parameters than those used in
the analysis of our previously reported $^{12}_{\Lambda}$B spectrum. 
The newer analysis reported here improved the missing mass
resolution, and more significantly, reduced the tails of the spectral peaks.

The simulation of charged particle trajectories through the system 
of magnets and detectors used the program, RAYTRACE\cite{raytrace}, 
and the coefficients of the RAYTRACE code were determined by 
adjusting the splitter contribution so that the calculated multi-dimensional, 
phase-space distributions from a point beam matched 
those measured when the entrance 
angles and positions of reaction protons and pions from the target 
were restricted by a set of appropriately positioned 
holes in a tungsten plate (sieve slit) located between the splitting 
magnet and the SOS\cite{tang}.  Optimization of the SOS 
coefficients used a $\chi^{2}$ minimization process
defined by the difference between the simulated and observed 
experimental patterns. 

The calibration procedure used an initial set of transport
parameters for the SOS-splitter spectrometer, and an initial focal plane 
calibration for the ENGE
spectrometer.  Then an adjustment to the beam energy and central
momentum of the spectrometers was used to simultaneously match 
the correct kinematic positions and widths of the
$\Lambda$ and $\Sigma$ peaks as produced in the $p(\gamma,K^{+})Y$
reaction, Figure \ref{fig:cal}.
The fit was also subject to the requirement that a reasonable 
position and width for the excitation of the ground state doublet was
obtained in the
$^{12}C(\gamma,K^{+})^{12}_{\Lambda}B$ reaction.  This later constraint
removed ambiguities in the global parameter space of
reconstruction coefficients, but did not place specific conditions on
the ground state splitting, excitation strength, or binding energy.
It only constrained these observables
to lie within an expected range of permissible values\cite{bayesian}.

After obtaining a consistent set of absolute beam energy and central
momentum values, the transport coefficients were re-fitted to the sieve
slit data and the Enge focal plane calibration was re-determined.
Residual mass offsets of $\le 200$ kev were obtained, and these were
dominated by statistics.  The peak widths were in good 
agreement with a Monte Carlo simulation having the same statistics 
and background levels as the data. 

From this more extensive calibration, the analysis produces 
peak widths of 2.8 MeV, 2.1 MeV and 0.75 Mev for the $\Lambda$ ,
$\Sigma$, and ground state doublet of $^{12}_{\Lambda}$B, respectively.
The previously reported values\cite{miyoshi1} were, 3.5 MeV, 2.7 MeV,
and 0.90 Mev. 
It is estimated that the new calibration procedure produces a 300 keV
error in the missing mass over the 130 MeV spectrometer acceptance.

The broad distribution above background and below the $\Lambda$ 
missing mass as seen in Figure \ref {fig:cal} was due to hyperon  
production from the $^{nat}$C in the target.  Unfortunately during the 
extended calibrations runs, the C to H ratio changed so that a direct  
normalization using the known $p(\gamma,K^{+})\Lambda$ cross section 
could not be applied.

\section{EXPERIMENTAL RESULTS AND DISCUSSION}

The experiment obtained data for both $^{12}_{\Lambda}$B and
$^{7}_{\Lambda}$He hypernuclei.  The
$^{12}_{\Lambda}$B spectrum was reported earlier.  Subsequently the
spectrometer transport and calibrations have been more extensively
studied, as discussed above, with the result that the 
experimental resolution is now
$\approx \, 750$ keV (FWHM).  The new spectrum is presented below, and in
addition, we also show for the first time our $^{7}_{\Lambda}$He results.

\subsection{ Spectroscopy of $^{12}_{\Lambda}$B Hypernuclei}

\begin{figure}[htb!]
\begin{center}
\epsfxsize = 7cm
\epsffile{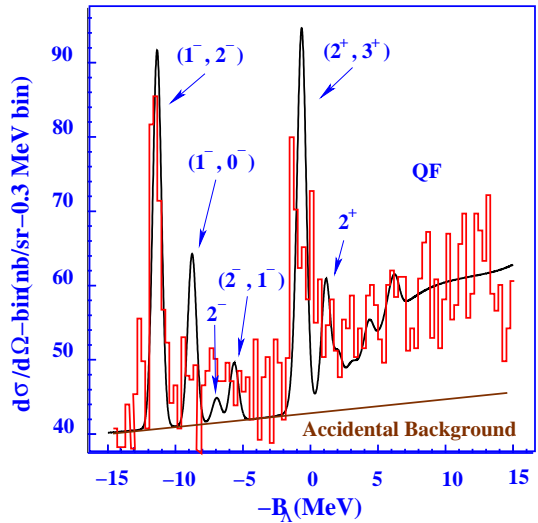}
\caption{The binding energy spectrum for $^{12}_{\Lambda}$B
  electroproduced from a $^{nat}$C target. The solid histogram
is the measured
accidental background, and the curve is a theoretical calculation,
spread by 750 keV and overlayed on, not fit to, the data.}
\label{fig:c-spectrum}
\end{center}
\end{figure}

\begin{figure}[htb!]
\begin{center}
\epsfxsize = 7cm
\epsffile{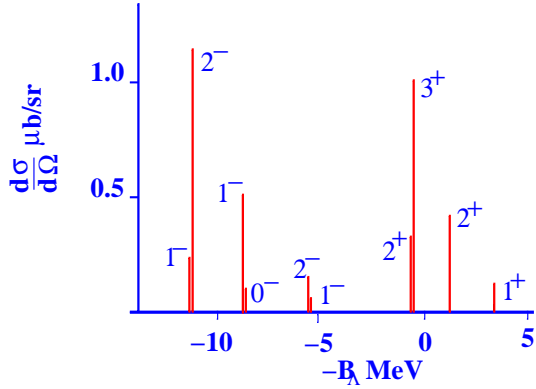}
\caption{A schematic representation of the reaction strength for the
  low lying states of $^{12}_{\Lambda}$B}
\label{fig:bL-reaction-strength}
\end{center}
\end{figure}

The binding energy spectrum with background of the $^{12}_{\Lambda}$B hypernucleus is
shown in Figure \ref{fig:c-spectrum}.
Two prominent structures are obvious in the spectrum.  The spectrum
is similar to 
that predicted by Motoba, {\it et al} \cite{motoba1,motoba2} and by
Millener\cite{millener3}.  
Reference \cite{motoba2} calculates
the excitation strengths in DWIA for the photoproduction process of a kaon
at an angle of $3^{\circ}$ by a 1.3 GeV photon.  Our original
publication compared the experimental spectrum to a calculation at
$0^{\circ}$ and a 1.2 GeV photon energy.
The curve in Figure \ref{fig:c-spectrum} is
generated by superimposing Gaussian peaks of the strength and at the
energy of each state as obtained from
a theoretical prediction.  For this superposition, the peak
widths are assumed to be 750 keV (FWHM) below and 5 MeV above 15 MeV 
excitation energy.  The background is obtained from 
a polynomial fit to
the averaged accidental background.  The positions of the states are
taken from ref 
\cite{millener3}, as this latter spectrum was obtained from
an effective p shell $\Lambda$-nucleus
interaction previously matched to ($\pi^{+},K^{+}$) 
data. The reaction strengths \cite{motoba2} for the low-lying states of
$^{12}_{\Lambda}$B are shown in Figure
\ref{fig:bL-reaction-strength}. This 
theoretical curve is directly 
overlayed on (not fitted to) the data. 

The major excitations are in good statistical agreement with theory 
both in position and strength. However, the core excited states,
predicted to lie between the major shell excitations, seem
to be more weakly excited than predicted. Statistics are not sufficient 
to discuss this region of the spectrum in detail.  In comparison to
the earlier, published spectrum \cite{miyoshi1} of this hypernucleus,
the resolution and shape of the ground state doublet is improved, and
the fluctuations in the core excited region reduced.

The differential cross section can be
calculated as if it were 
photoproduction, by assuming the virtual photons are massless.
This averages the elementary ($\gamma$,K) reaction at
1500 MeV over the $\approx$
100 MeV spread of virtual gamma energies.   The weighted average of
the cross section measurements for the ground state doublet 
at the two incident beam energies 
is, $140 \, \pm \, 17(\mbox{stat}) \, \pm \, 18(\mbox{sys})$ 
nb/sr.  This value is consistent with the individual values of the
separate energy measurements, and also the theoretical 
photoproduction prediction for this angle and energy \cite{motoba2}, 
$\approx$ 138 nb/sr.
 
The $2^{-}$ component of the ground state doublet is predicted to lie at
approximately 150 keV excitation energy, and is expected to be
dominant, Figure ~\ref{fig:bL-reaction-strength}.  
The resolution (and statistics) is insufficient to identify
this splitting.  The $3^{+}$  p-shell state is also predicted to
dominate the spectrum in the $\approx \, 10$ MeV excitation region.  
While theory indicates the dominant structure at this excitation energy
is actually created by the overlap, Figure ~\ref{fig:bL-reaction-strength}, 
of the $2^{+}$ and $3^{+}$ p-shell states\cite{millener3}. 
These two states are not as degenerate in other calculated spectra
\cite{motoba2} which use
slightly different effective parameters. The newer data
suggest that this p-shell peak may be broader than first indicated,
but statistics limit the ability to draw specific conclusions.
The results do demonstrate sensitivity to the effective 
interaction and the DWIA transition amplitudes.

The binding energy scale is determined from the position of the
$\Lambda$ and $\Sigma$ peaks in the calibration spectrum. The
$^{12}_{\Lambda}$B binding energy is found to be $11.52 \, \pm \,
0.35$ MeV and is in agreement with the accepted value \cite{juric}
obtained from a measurement using emulsion, 11.37 MeV.  

To confirm our cross section normalization, the quasi-free component 
of the experimental spectrum was extracted, 
and the yield corrected for acceptance and momentum transfer
\cite{bebek}. We obtained 4.2 interacting
protons, in agreement with previous measurements \cite{hinton,maeda}.
 
\subsection{ Spectroscopy of $^{7}_{\Lambda}$He Hypernuclei}

\begin{figure}[htb!]
\begin{center}
\epsfxsize = 15cm
\epsffile{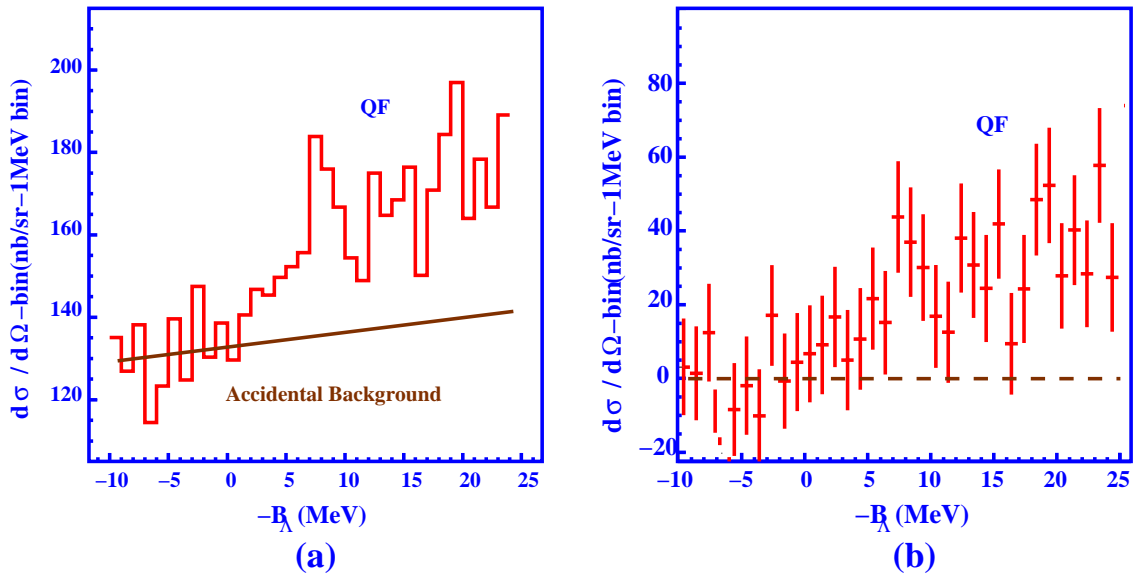}
\caption{The binding energy spectrum for $^{7}_{\Lambda}$He
  electroproduced from a $^{7}$Li target. The data are binned in 1 MeV
  intervals.  In figure (a) the zero in the cross section scale is
  suppressed. Figure (b)  shows the background subtracted 
spectrum with statistical errors. }
\label{fig:he7L_comb}
\end{center}
\end{figure}

\begin{figure}[htb!]
\begin{center}
\epsfxsize = 7 cm
\epsffile{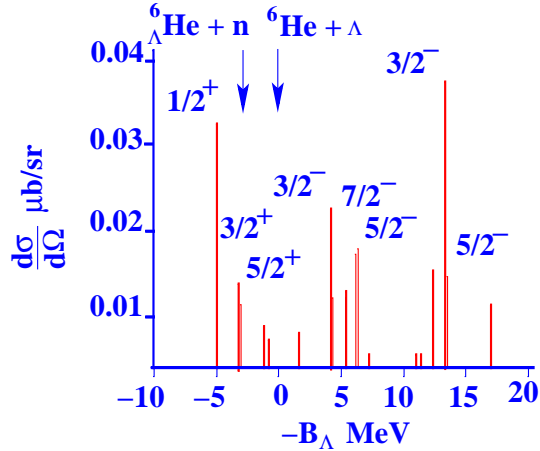}
\caption{A schematic representation of the reaction strength for the
  low lying states of $^{7}_{\Lambda}$He.}
\label{fig:heL-reaction-strength}
\end{center}
\end{figure}

The measured $^{7}_{\Lambda}$He spectrum from the 
$^{7}$Li$(e,e^{\prime}K^{+})^{7}_{\Lambda}$He reaction 
with the accidental background 
is shown in Figure \ref{fig:he7L_comb}. Data are binned in 1 MeV
intervals to improve statistics, although the 
experimental resolution is comparable to that obtained in
the $^{12}_{\Lambda}$B spectrum (750 keV). The threshold ($B_{\Lambda}
= 0$) is defined as the energy 
between an unbound $\Lambda$ and 
a $^{6}$He core.  Negative energies represent bound $\Lambda$
states, but in this case the $^{6}_{\Lambda}$He + n and
$^{5}_{\Lambda}$He + (nn) 
thresholds are some $\approx$ 2 MeV below the threshold value as defined above.
The theoretical prediction\cite{richter} of the reaction strengths 
is shown in Figure ~\ref{fig:heL-reaction-strength}.
There is little resemblance to the data,
although the experimental statistics are poor.  For example there is
no evidence of any bound state excitations, although the ground state 
has a predicted strength of 30 nb/sr.  
Statistics limit the interpretation of the
spectrum as the average statistical error is about 13 nb/sr-MeV between
-10 and 0 MeV.  Using the expected resolution, the total statistical
error in the data is approximately 16 nb/sr summed over a 4
$\sigma$ width about the expected ground state position.  As discussed
in ref. \cite{hiyama,richter}, the cluster structure of the
$^{7}_{\Lambda}$ hypernuclear system will effect the intrinsic widths of
the excited states.

The data suggest a peak at about 7 MeV above the
$\Lambda$-$^{6}$He threshold, having a statistical uncertainty of about
9\% and a width of approximately 3 Mev (FWHM). The background
subtracted spectrum with statistical error is shown in Figure
~\ref{fig:he7L_comb}.  If the enhancement at $\approx \, 7 MeV$ in the
data corresponds to the
superposition of the predicted states near 5 MeV and is not a statistical
fluctuation, then
the energies are incorrect and the strength is somewhat less than
predicted. However the widths of these individual states must be due 
to their intrinsic values, which are broader than the experimental 
resolution.
Except for the tail of the quasi-free spectrum, no significant features are 
observed in the data.

The $^{7}_{\Lambda}$He system has
two loosely-bound, p-shell neutrons added to a 
$^{4}$He core.  The addition of the $\Lambda$ should significantly
perturb the nuclear core, shrinking the nuclear radius.  Such a perturbation is
observed in the $^{7}_{\Lambda}$Li hypernuclear system as a change in the
$B(E2)$ gamma transition rate between the $5/2$ and $1/2$
states\cite{tanida}.  
Thus it is expected to find a $^{7}_{\Lambda}$He hypernuclear
state\cite{hiyama} 
bound by about 5 MeV. The ground state masses of the light hypernuclei 
(s and p shell) are generally 
determined by the emulsion experiments.  In the case of this
hypernucleus, the mass distribution
from the various emulsion experiments was so broad that a consistent binding 
energy could not
be determined.  Isomeric states could 
explain this broad width\cite{gal}, as 
the binding energy is obtained from the energies of 
the decay products.  However this would 
not effect the widths of the reaction peaks in our
experiment.  Clearly more experimental investigations are 
needed, as well as better treatments of the structure of $^{7}_{\Lambda}$He
and the $^{7}$Li$(e,e^{\prime}K^{+})^{7}_{\Lambda}$He
reaction mechanism. 

\section{CONCLUSION}

The first electroproduction experiment using the high quality 
electron beam at JLab demonstrated a sub-MeV energy resolution.
The experiment was carefully crafted to optimize the resolution and
rates using the existing SOS magnetic spectrometer.
The $^{12}_{\Lambda}$B spectrum was similar to that predicted by theory
both in the position and magnitude of the major excitations.  However
the spectrum of $^{7}_{\Lambda}$He was not well reproduced by the 
existing reaction calculation.

High resolution, systematic studies 
of electroproduced spectra can
complement hypernuclear studies by hadronic probes and gamma spectroscopy. 
The high-quality electron beam at JLab provides new opportunities for future 
hypernuclear studies with better resolution and much better
quality.  A new experiment based on the experience obtained from 
this experiment is expected to
result in resolutions improved by a factor of 2 over the value
reported here, with rates increased by
a factor of 40.  A newly designed Kaon spectrometer  will be
dedicated to this research \cite{e01_011}.

	In summary, the investigation of strangeness in nuclear systems is not
merely an extension of conventional nuclear physics.  Certainly one
cannot, nor would one want to, reproduce the wealth of information
that has been accumulated on conventional nuclei.  The hypernucleus,
however, offers a selective probe of the hadronic many-body problem,
providing insight in areas that cannot be easily addressed by
other techniques.  What is now needed is a
series of precision studies with high resolution where
level positions and weak decay dynamics can be compared to theory. 


\begin{acknowledgements}

Support of the Accelerator and Physics Division staff 
of the Thomas Jefferson National Accelerator Facility (JLab) is
gratefully acknowledged. We wish
to thank Dr. D.  
J. Millener and Dr. T. Motoba for many useful discussions.  The Southeastern 
University Research Association (SURA) operates JLab for the
U.S. Department of  
Energy under Contract No. DE-AC05-84ER40150. This work is also
supported in part by 
research grants from the U.S. Department of Energy and the National Science 
Foundation, by Monkasho Grant-in-aid for Scientific Research
(09304028, 09554007, 12002001), and by the US-Japan collaborative
research program under the auspices of the NSF and Japan Society for
Promotion of Science.

\end{acknowledgements}


\begin{thebibliography}{99}
\bibitem{bayesian}Although Bayesian probability analysis was not
  directly used here, the application of prior information on allowed
  values of a  likelihood were used in this spirit.
\end{thebibliography}
\end{document}